\documentclass{article}
\usepackage{ijcai25}

\usepackage{times}
\usepackage{soul}
\usepackage{url}
\usepackage[hidelinks]{hyperref}
\usepackage[utf8]{inputenc}
\usepackage[small]{caption}
\usepackage{graphicx}
\usepackage{amsmath}
\usepackage{amsthm}
\usepackage{booktabs}
\usepackage{algorithm}
\usepackage{algorithmic}
\usepackage[switch]{lineno}

\title{Fresh2comm: Information Freshness Optimized Collaborative Perception}

\author{
Ziyong Wu$^1$
\and
Zhilin Peng$^2$\and
Lei Yu$^*$
\affiliations
Sino-French Engineer School, Beihang University, Beijing, China
\emails
jimfor6@163.com,
pengzlva@163.com,
yulei@buaa.edu.cn
}

\begin{document}

\maketitle

\begin{abstract}
    Collaborative perception is a cornerstone of intelligent connected vehicles, enabling them to share and integrate sensory data to enhance situational awareness. However, measuring the impact of high transmission delay and inconsistent delay on collaborative perception in real communication scenarios, as well as improving the effectiveness of collaborative perception under such conditions, remain significant challenges in the field. To address these challenges, we incorporate the key factor of information freshness into the collaborative perception mechanism and develop a model that systematically measures and analyzes the impacts of real-world communication on collaborative perception performance. This provides a new perspective for accurately evaluating and optimizing collaborative perception performance. We propose and validate an Age of Information (AoI)-based optimization framework that strategically allocates communication resources to effectively control the system's AoI, thereby significantly enhancing the freshness of information transmission and the accuracy of perception. Additionally, we introduce a novel experimental approach that comprehensively assesses the varying impacts of different types of delay on perception results, offering valuable insights for perception performance optimization under real-world communication scenarios.

\end{abstract}

\section{Introduction}

Intelligent Connected Vehicles (ICV) represent a profound integration of automotive and communication technologies, emerging as a pivotal direction for the future development of the automotive industry. Collaborative perception, a core functionality of ICV, leverages advanced communication and computing technologies to enable vehicles to share sensory data. This process not only expands the perception range but also enhances the accuracy of situational awareness. By overcoming the limitations of individual vehicle sensors, collaborative perception significantly improves driving safety.
Recent state-of-the-art (SOTA) studies, such as Where2comm \cite{NEURIPS2022_1f5c5cd0} and How2comm \cite{NEURIPS2023_4f31327e}, have achieved notable progress in this domain. These studies primarily focus on improving the accuracy of collaborative perception under the assumption of constant information transfer latency. However, in real-world communication scenarios of ICV, significant uncertainty and inconsistency in information transmission delay between vehicles arise due to the rapid movement of vehicles and interference in vehicle-to-vehicle communication. This inconsistency in information transmission delay further affects the accuracy of collaborative perception.
Simulations of ICV communication reveal that the information transmission delay between vehicles may last a long time, with the delay variation range extending from a few milliseconds to several seconds. Consequently, quantifying the impact of high and inconsistent delay on collaborative perception in real communication scenarios, and enhancing the effectiveness of collaborative perception under conditions of high delay and significant inconsistency, are critical challenges that must be addressed in contemporary collaborative perception research.

In contemporary research, the ICV system is modeled as a multi-agent system, where vehicles function as autonomous agents, and the collaborative perception results are generated through interactions among these agents \cite{NEURIPS2023_4f31327e}. Consequently, information delay in this context is not solely attributable to transmission delay but also encompasses processing delay. The Age of Information (AoI) is defined as the time elapsed since the most recent data packet was received \cite{6195689}. In the realm of collaborative perception, AoI thus provides a holistic measure of information freshness. 
To comprehensively assess the impact of information delay on collaborative perception outcomes, it is imperative to account for both computation and communication delays. An analysis of their respective influences can elucidate potential optimization directions, thereby providing insights into reducing collaborative perception errors within complex communication scenarios.

To mitigate the adverse effects of information delay on collaborative perception, we initially focus on optimizing the system's information transmission delay and addressing the inconsistency of transmission delays without imposing a significant computational burden on vehicles. To this end, we formulate optimization objectives that aim to minimize the maximum information transmission delay. This approach not only reduces the overall system transmission delay but also effectively mitigates the issue of inconsistent delays.
Traditional optimization methods, however, often entail high computational complexity, which may not be suitable for resource-constrained vehicular systems. To circumvent this limitation, we propose a greedy-based optimization solver. This solver efficiently optimizes information transmission delay with minimal additional computational load on vehicles, thereby ensuring the optimal overall AoI of the system.

The contributions of this work are summarized as follows:
\begin{itemize}
  \item We integrate the critical factor of information freshness into the collaborative perception mechanism, thereby developing a model that closely reflects real-world communication scenarios. This integration provides a novel perspective for accurately evaluating and optimizing collaborative perception performance.
  \item We propose and validate an AoI-based optimization framework that strategically allocates communication resources, effectively controlling the system's AoI. This significantly enhances the freshness of information transmission and the accuracy of perception.
  \item We introduce a novel experimental approach that comprehensively assesses the varying impacts of different types of delay on perception results. This approach offers valuable insights for multi-dimensional optimization of collaborative perception systems.
\end{itemize}

\section{Related works}
Collaborative perception is a foundational technology in ICV, enabling vehicles to share and fuse sensory data to enhance situational awareness. Despite significant advancements, challenges such as data asynchrony, communication efficiency, and the impact of information freshness persist, necessitating further investigation.

Substantial progress has been achieved in addressing data sharing and fusion challenges within collaborative perception. The study by \cite{hu2024collaborativeperceptionconnectedautonomous} identifies critical issues, including data asynchrony, and proposes a dynamic communication graph framework to minimize latency. Similarly, the work in \cite{DBLP:journals/sensors/CuiZXYF22} emphasizes the role of wireless communication in enhancing perception accuracy through environmental information fusion. Furthermore, edge-assisted frameworks such as  \cite{9709015} demonstrate how V2X communication can effectively improve environmental awareness in dynamic vehicular networks.

Optimizing communication efficiency remains a vital focus for collaborative perception, particularly in bandwidth-constrained environments. The Where2comm framework proposed in \cite{NEURIPS2022_1f5c5cd0}, reduces bandwidth consumption by prioritizing spatially significant data for transmission. Complementing this, the How2comm framework in \cite{NEURIPS2023_4f31327e} utilizes mutual information-based mechanisms and delay compensation strategies to optimize multi-agent collaboration. Additionally, frameworks like What2comm \cite{10.1145/3581783.3611699} and When2com \cite{9156848} dynamically balance communication overhead and perception accuracy through innovative feature selection and communication graph grouping methods.

Publicly available datasets play an essential role in advancing collaborative perception research by providing standardized benchmarks to evaluate fusion strategies and real-world performance. For instance, the OPV2V dataset introduced by \cite{10.1109/ICRA46639.2022.9812038} , supports V2V communication research with over 11,000 frames and multiple fusion pipelines. Similarly, the DAIR-V2X dataset proposed by \cite{9879243} enables studies on asynchronous fusion and communication delays using diverse and synchronized scenarios. These datasets are instrumental in addressing challenges such as communication variability and the alignment of multi-source data.

The concept of AoI has emerged as a critical metric for maintaining information freshness in collaborative perception systems. The work in  \cite{9903381} explores the interplay between AoI, latency, and reliability, emphasizing the impact of dynamic vehicular conditions on AoI variations. An AoI-driven power allocation strategy is proposed in \cite{10081395}, balancing AoI reduction with system throughput to ensure accurate and timely information updates. In edge-enabled environments, the study \cite{9599398} addresses the dual challenges of AoI and computational delays, proposing efficient resource allocation methods to maintain information freshness. Additionally, \cite{8954939} employs deep reinforcement learning to optimize wireless resource allocation and minimize AoI in dynamic vehicular networks.  Moreover, the study by \cite{8937801} focuses on minimizing the tail distribution of AoI, targeting ultra-reliable, low-latency vehicular communications. These studies collectively demonstrate the critical role of AoI in enhancing the timeliness and accuracy of collaborative perception.

To further enhance the performance of collaborative perception, several studies investigate the joint optimization of communication and computation processes. The active learning-based approach in \cite{8918241} dynamically allocates resources to minimize AoI while ensuring communication reliability. Similarly, the work in \cite{9023385} proposes joint strategies for optimizing AoI and completion time, establishing a theoretical foundation for balancing timeliness and efficiency. Finally, the study \cite{10130737} explores the integration of trajectory planning and resource allocation in UAV-assisted vehicular networks, demonstrating the potential of multi-modal collaboration to enhance information freshness and perception performance.

\section{Model}

\begin{figure}[htbp]
    \centering
    \includegraphics[width=0.99\linewidth]{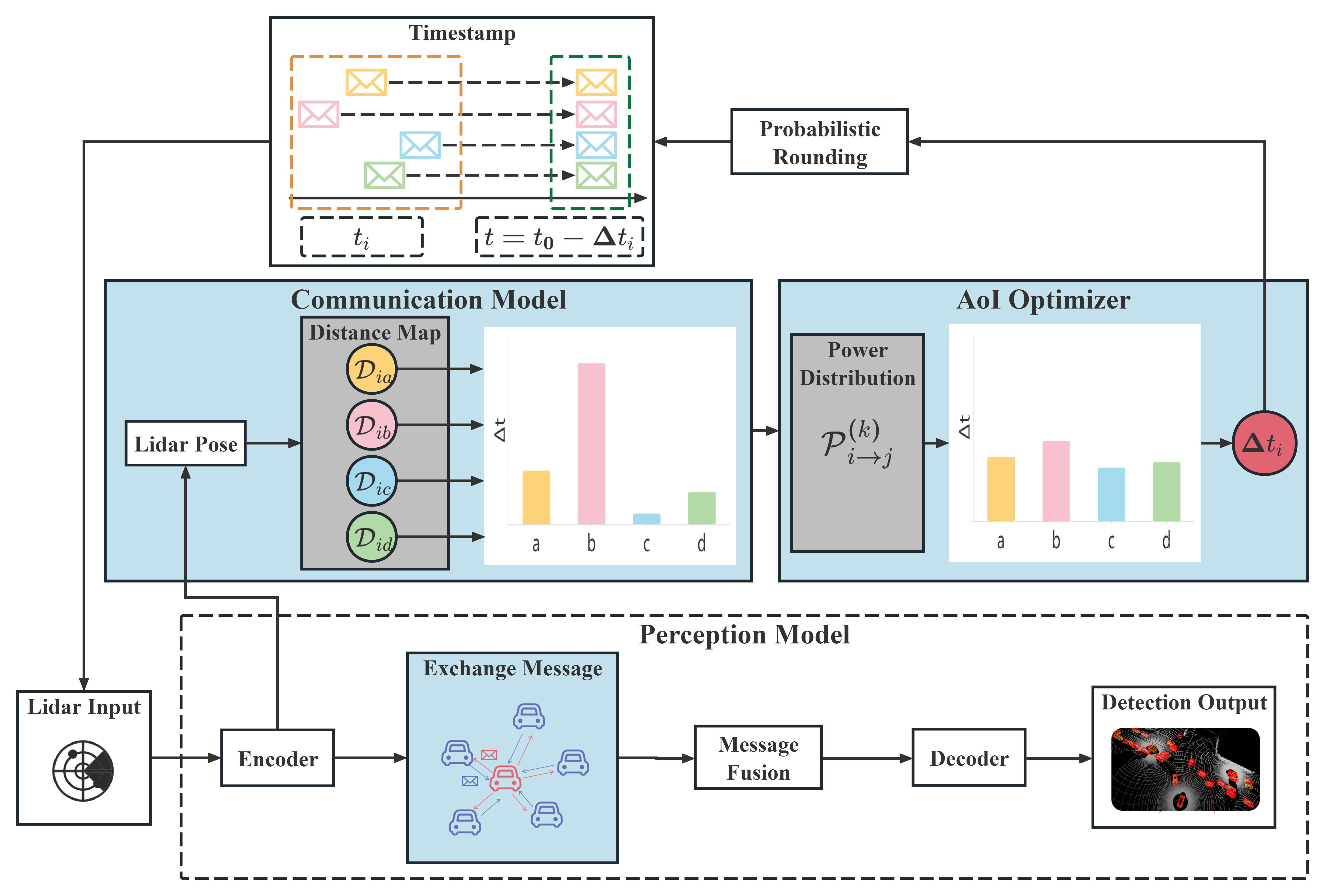}
    \caption{System overview}
    \label{fig:enter-label1}
\end{figure}

This section presents \textit{Fresh2comm}, an integrated collaborative perception framework that incorporates real-world communication models and AoI optimization. An overview of the framework is depicted in Figure~\ref{fig:enter-label1}. \textit{Fresh2comm} comprises three primary modules: a perception model, a communication model, and an AoI optimizer. Specifically, the communication model calculates the communication delays between vehicles based on a distance map, thereby accurately reflecting real-world communication scenarios. Simultaneously, the AoI optimizer regulates the system's AoI by managing both computation delay growth and communication delay control.

\subsection{Perception Architecture}

 As illustrated in Figure~\ref{fig:enter-label1}, an encoder is employed to obtain the LiDAR pose, which is subsequently projected into a distance map representing the distances between the communication vehicles. The communication delay is calculated based on a real-world communication model, while the computational delay associated with the perception process is determined through experimental measurement.

Following the AoI optimization step, the total delay is adjusted using \textit{Probabilistic Rounding} and returned with the corresponding timestamp of data. Subsequently, the original LiDAR data are retrieved based on the calculated timestamp and utilized for the perception process. This approach effectively simulates the degradation of LiDAR data freshness due to communication and computational delays, thereby replicating realistic operational conditions in a real-world scenario.

\subsection{Communication Model}

Consider a scenario where each vehicle communicates with others via a V2V (Vehicle-to-Vehicle) communication network, in which $n$ vehicles share a common channel.

The objective is to calculate the transmission delay between any two vehicles $i$ and $j$. In this scenario, the underlying network is based on V2V communication. To simplify the analysis, we model the communication interference by categorizing it into three primary components: inter-signal interference, signal fading, and environmental noise \cite{9788527}.

Specifically, inter-signal interference arises from the competition for transmission resources between different signals on the same channel. In a multi-vehicle communication scenario, communication signals between vehicles interfere with each other, affecting the quality of data transmission. Signal fading refers to the weakening of the signal strength due to factors such as path loss, shadowing effects, and multipath propagation. Lastly, environmental noise represents the interference caused by various random noise sources in the wireless communication environment, such as thermal noise and electromagnetic interference, which are inherent in any communication system.

Finally, the transmission delay $ \text{Delay}_{ij} $ and the signal-to-noise ratio $ \text{SNR}_{ij} $ between vehicles $i$  and $ j $ can be formulated as follows:
\begin{equation}
    \text{Delay}_{ij} = \frac{S}{B \log_2 \left( 1 + \text{SNR}_{ij} \right)}
\end{equation}

\begin{equation}
    \label{equa:snr}
    \text{SNR}_{ij} = \frac{P_{ij}} {{D_{ij}^{\alpha}} * {(\sum_{k \neq i} \frac{P_{kj}}{D_{kj}^{\alpha}}+n)}}
\end{equation}

where:

\begin{itemize}
    \item $ \text{Delay}_{ij} $ denotes the transmission delay from vehicle $ i $ to vehicle $ j $,
    \item $ \text{SNR}_{ij} $ represents the signal-to-noise ratio between vehicle $ i $ and vehicle $ j $, 
    \item $ B $ is the communication bandwidth,
    \item $ S $ is the transmitted bits for communication,
    \item $ P_{ij} $ is the transmit power of vehicle $ i $ to vehicle $ j $,
    \item $ D_{ij} $ is the distance between vehicle $ i $ and vehicle $ j $,
    \item $ \alpha $ is the path loss exponent,
    \item $ n $ is the noise power.
\end{itemize}

This formula accounts for the direct power from vehicle $ i $ to vehicle $ j $ and the interference from all other vehicles. The resulting SNR matrix $ \mathbf{SNR} $ is an $ n \times n $ matrix, where each entry $ \text{SNR}_{ij} $ represents the SNR from vehicle $i$ to vehicle $j$.

\subsection{AoI Optimizer}
\subsubsection{Problem Formulation}
The objective of the power allocation problem is to maximize the minimum SNR across the network while adhering to power constraints. The optimization problem can be formulated as follows:
\begin{equation*} 
\centering
    \min_{\mathbf{P}} \max_{i,j} \text{Delay}_{ij}
\end{equation*}
or as well:
\begin{equation*} 
\centering
    \max_{\mathbf{P}} \min_{i,j} \text{SNR}_{ij}
\end{equation*}
subject to the following constraints:

\begin{itemize}
    \item The transmit power $ P_{ij} $ of each vehicle $ i $ to vehicle $ j $ must satisfy:\\
    \begin{equation*} 
    \centering
    P_{\text{min}} \leq P_{ij} \leq P_{\text{max}}, \quad \forall i,j
    \end{equation*}
    \\
    where $ P_{\text{min}} $ and $ P_{\text{max}} $ is the limit of transmit power.
    
    \item The total power transmitted by each vehicle must not exceed the maximum allowed power $ P_{\text{max}} $:
    \begin{equation*} 
    \centering
    \sum_{j \neq i} P_{ij} \leq P_{\text{max}}, \quad \forall i
    \end{equation*}

\end{itemize}
\subsubsection{GreedyPA Method}
As previously discussed, achieving effective control and optimization of the AoI necessitates the management of communication delays while ensuring low computational complexity in the algorithms employed. To address this requirement, we have developed a lightweight greedy algorithm that, while satisfying power constraints, maximizes the minimum SNR and maintains low algorithmic complexity. The proposed greedy method is detailed in Algorithm~\ref{alg:greed}.

\begin{algorithm}[htbp]
    \caption{Greedy Algorithm-based Power Allocation}
    \label{alg:greed}
    \textbf{Input}: distance matrix, number of vehicles\\
    \textbf{Parameter}: learn rate, maximum epoch\\
    \textbf{Output}: optimized power matrix
    \begin{algorithmic}[1] 
        \STATE Initialize the power matrix.
        \FOR{epoch = 1 to max\_epochs}
            \STATE Calculate max\_SNR and min\_SNR according to Equation~\ref{equa:snr}
            \STATE Adjust $ P(\text{max\_SNR}) $ and $ P(\text{min\_SNR}) $ using the adaptive method.
            \STATE Reconstruct the $ P $ matrix to satisfy the constraints.
            \IF{enough optimized}
                \STATE break.
            \ENDIF
        \ENDFOR
        \STATE \textbf{return} solution found
    \end{algorithmic}
\end{algorithm}

\section{Evaluation}

We designed and implemented a series of simulation experiments. The parameter selection for our communication model refers to \cite{9023385} and \cite{NEURIPS2023_4f31327e}, as well as some non-paper data. The key parameters are listed in Table~\ref{tab:parameter_settings}:

\begin{table}[htbp]
    \centering
    \begin{tabular}{cc}
        \toprule
        \textbf{Parameter} & \textbf{Value} \\
        \midrule
        Alpha  & $3$ \\
        Width & $10 \, \text{MHz}$ \\
        Base Noise  & $4.14 \times 10^{-14} \, \text{W}$ \\
        Maximum Power  & $23 \, \text{W}$ \\
        Minimum Power  & $10^{-6} \, \text{W}$ \\
        Data Size  & $1.06 \, \text{MB}$ \\
        \bottomrule
    \end{tabular}
    \caption{Parameter Settings}
    \label{tab:parameter_settings}
\end{table}

\subsection{Optimization Method Evaluation}
In this section, we compare three optimization strategies for delay reduction: \textit{Greedy Algorithm-based Power Allocation (GreedyPA), Default Power Allocation (DefaultPA), Genetic Algorithm-based Power Allocation (GeneticPA)}, and we focus on evaluating the performance of the designed GreedyPA. GeneticPA, which is developed based on the Genetic Algorithm (GA) [Lambora et al., 2019], serves as a benchmark for evaluating the accuracy of GreedyPA in solving the AoI optimization problem. Additionally, DefaultPA, which uniformly distributes the transmit power among vehicles, is employed to compare the effectiveness of GreedyPA.

\subsubsection{Baseline: GeneticPA}
In Algorithm~\ref{alg:genetic}, the population size is set to 50, with a crossover rate of 0.8 and a mutation rate of 0.05. The algorithm evolves for a maximum of 100000 generations, with each generation selecting the best candidates based on their fitness values.

\begin{algorithm}[htbp]
    \caption{Genetic Algorithm-based Power Allocation}
    \label{alg:genetic}
    \textbf{Input}: Initial population, distance matrix, number of vehicles\\
    \textbf{Parameter}: Population size, crossover rate, mutation rate, maximum generations\\
    \textbf{Output}: Optimized delay matrix
    \begin{algorithmic}[1] 
        \STATE Initialize population with random power allocations.
        \FOR{generation = 1 to max\_generations}
            \FOR{each individual in population}
                \STATE Calculate fitness based on minimum SNR values.
                \IF{fitness is below threshold}
                    \STATE Discard individual and select a new one.
                \ENDIF
            \ENDFOR
            \STATE Select parents based on fitness.
            \STATE Perform crossover and mutation.
            \STATE Generate new population.
        \ENDFOR
        \STATE \textbf{return} best solution found
    \end{algorithmic}
\end{algorithm}

\subsubsection{Experimental Setup}
To evaluate the effectiveness of our delay optimization methods in real-world scenarios, we initially employed the DAIR-V2X dataset \cite{9879243}, the world’s first large-scale, multimodal, and multi-perspective dataset for vehicle-road collaborative autonomous driving research. We chose to construct a distance matrix using the actual positions of vehicles in the annotated scenes from the dataset. Additionally, based on the vehicle density in these scenes, we estimate that within a certain range, a single channel can support communication among up to approximately 5 vehicles. Therefore, we simulate the single-channel communication scenarios for 3 vehicles, 4 vehicles, and 5 vehicles, respectively. 

Overall, in the annotated scenes of DAIR-V2X, we compared three optimization strategies:

\begin{itemize}
    \item \textbf{DefaultPA}: The baseline approach where the power is evenly allocated: $
    P_{ij} = P_{\text{max}}/(n-1),\quad \forall i,j
    $
    \item \textbf{GreedyPA}: A lightweight heuristic optimization method proposed to minimize the maximum communication delay.
    \item \textbf{GeneticPA}: The baseline method that simulates the optimal channel gain allocation, used to compare the delay results from the other two methods.
\end{itemize}

To evaluate the overall performance improvements of each strategy, we analyzed and compared the communication delays obtained from DefaultPA and GreedyPA, with the delay results derived from GeneticPA. The difference in delays was calculated using the RMSE (Root Mean Square Error) method. Additionally, we compared the variance and the average value in communication delays among the three methods. The variance in communication delays can assess whether any of the channel gain allocation strategies would lead to significant information asynchrony issues. We randomly selected scenes and vehicle counts from the annotated dataset, performing 15 tests for single-channel communication involving 3, 4, and 5 vehicles, respectively, and then calculated the average RMSE, variance and average value. These results provide valuable insights into the effectiveness of our methods in mitigating delay-related challenges in collaborative perception.

\subsubsection{Evaluation Result}
As shown in Table~\ref{tab:optimization_comparison}, we conducted a series of experiments. First, by comparing DefaultPA with GreedyPA, we observed a significant gap between the communication delay under DefaultPA and the approximately optimal solution derived from GeneticPA. However, the results obtained from the GreedyPA were very close to the approximate optimal solution. Despite this, our designed algorithm still has room for optimization of several tens of milliseconds when the number of vehicles is large. 

In addition, we conducted a small-scale ablation study on the number of iterations as a hyper-parameter for GreedyPA. The results demonstrated that this Algorithm converges extremely quickly and, with its low time complexity, can effectively meet the low-latency requirements in real-world scenarios. 
\begin{table}[htbp]
    \centering
    \resizebox{8.5cm}{!}{ 
    \setlength{\tabcolsep}{4pt} 
    \begin{tabular}{cccc}
        \toprule
        \textbf{Evaluation Metric} & \textbf{cav=3} & \textbf{cav=4} & \textbf{cav=5} \\
        \midrule
        RMSE\_DefaultPA  & $26$ & $15$ & $79$ \\
        RMSE\_GreedyPA\_epoch5000  & $0.006$ & $0.02$ & $0.06$ \\
        RMSE\_GreedyPA\_epoch500  & $0.009$ & $0.03$ & $0.06$ \\
        RMSE\_GreedyPA\_epoch50  & $0.03$ & $0.09$ & $0.1$ \\
        \midrule
        VAR\_DefaultPA  & 1600 & $409$ & $ 3 \times 10^4 $ \\
        VAR\_GreedyPA  & $ 1 \times 10^{-7} $ & $0.003$ & $0.007$  \\
        VAR\_GeneticPA  & $ 5 \times 10^{-5} $ & $ 1 \times 10^{-5} $ & $0.002$ \\
        \midrule
        MEAN\_DefaultPA  & $15.2$ & $7.34$ & $44.3$ \\
        MEAN\_GreedyPA  & $0.107$ & $0.184$ & $0.258$  \\
        MEAN\_GeneticPA  & $0.106$ & $0.176$ & $0.251$ \\
        \bottomrule
    \end{tabular}
    }
    \caption{Optimization Strategy Comparison}
    \label{tab:optimization_comparison}
\end{table}

On the other hand, with respect to the variance and average value in delay, GreedyPA outperforms DefaultPA, achieving a very low value, comparable to the approximate optimal solution from GeneticPA. 

In summary, our GreedyPA significantly improves the averaging and minimization of delay compared to the default setting. It achieves near-optimal results while maintaining low time complexity.

\subsection{Evaluation of Perception Model under Different Delays}
\subsubsection{Experimental Setup}
In this section, we apply different types of delays to a pre-trained perception model and analyze the resulting perception outcomes. The perception framework we adopt is Where2comm. According to the definition of AoI, delays in collaborative perception arise from both computation and communication delays. For testing purposes, we separately examine the impact of three types of delays on the perception results:

\begin{itemize}
  \item \textbf{Computation Delay}: Simulates constant computation delays, and is set to all cavs.
  \item \textbf{Communication Delay}: Simulates fixed transmission delays, and is set to all other cavs.
  \item \textbf{Linear Communication Delay}: Simulates the asynchronous nature of transmission delays, is set as proportional to the vehicle distance.
\end{itemize}

Additionally, although the Dair-v2x dataset originates from real-world scenarios, it lacks perception data from multiple vehicles. Therefore, we use the OPV2V dataset for simulating collaborative perception \cite{10.1109/ICRA46639.2022.9812038}. 

\subsubsection{Experimental Result}
We organized the experimental results into Table~\ref{tab:delay_perception_part1}. For the evaluation of perception performance, we used three quantification dimensions: $AP@0.3$, $AP@0.5$, and $AP@0.7$. Generally, $AP@0.3$ is more focused on assessing whether perception has lost information, while $AP@0.7$ reflects the accuracy of the perception's positioning. 
\begin{table}[htbp]
    \centering
    \resizebox{8.5cm}{!}{
    \begin{tabular}{cccccc}
        \toprule
        \textbf{delay type} & \textbf{value} & \textbf{AP@0.3$\uparrow$} & \textbf{AP@0.5$\uparrow$} & \textbf{AP@0.7$\uparrow$} \\
        \midrule
        backbone(s)  & $0$  & $0.864$ & $0.859$ & $0.805$ \\
        backbone(s)  & $0.1$  & $0.855$ & $0.709$ & $0.148$ \\
        backbone(s)  & $0.2$  & $0.618$ & $0.183$ & $0.036$ \\
        backbone(s)  & $0.3$  & $0.258$ & $0.071$ & $0.021$ \\
        backbone(s)  & $0.4$  & $0.124$ & $0.045$ & $0.018$ \\
        backbone(s)  & $0.5$  & $0.081$ & $0.033$ & $0.017$ \\
        backbone(s)  & $1$ & $0.039$ & $0.024$ & $0.015$ \\
        \midrule
        trans\_delay(s)  & $0$  & $0.864$ & $0.859$ & $0.805$ \\
        trans\_delay(s)  & $0.1$  & $0.860$ & $0.810$ & $0.435$ \\
        trans\_delay(s)  & $0.2$  & $0.750$ & $0.481$ & $0.227$ \\
        trans\_delay(s)  & $0.3$  & $0.500$ & $0.332$ & $0.196$ \\
        \midrule
        liner\_coef(s/m)  & $0$  & $0.864$ & $0.859$ & $0.805$ \\
        liner\_coef(s/m)  & $0.1$ & $0.863$ & $0.836$ & $0.735$ \\
        liner\_coef(s/m)  & $0.5$ & $0.643$ & $0.440$ & $0.253$ \\
        liner\_coef(s/m)  & $1$   & $0.395$ & $0.314$ & $0.211$ \\
        \bottomrule
    \end{tabular}
    }
    \caption{Perception Results for Different Delay Settings}
    \label{tab:delay_perception_part1}
\end{table}

Specifically, when the delay type is \textit{backbone}, the performance degradation is the most rapid, with a noticeable nonlinear decline: the value of $AP@0.7$ declines much earlier and faster than $AP@0.3$. This suggests that backbone delays cause a shift in the overall perception results; within a certain range, information is not lost, but accuracy decreases. 

The second most significant type of delay is the constant transmission delay. Similar to backbone delays, its impact on perception performance is relatively weaker, but it still exhibits a clear nonlinear trend. 

In contrast, the distance-dependent linear delay experiment shows a marked reduction in nonlinearity, indicating that time asynchrony leads to unpredictable losses in perception results. Comparing $trans\_delay@0.2$ and $liner\_coef@1$, both exhibit similar $AP@0.7$ scores, but the $AP@0.3$ score for $liner\_coef@1$ is significantly lower than that of $trans\_delay@0.2$. This indicates that time asynchrony causes more critical information to be lost in perception.

In summary, both backbone delay and transmission delay affect perception accuracy to varying degrees, while the inconsistency in transmission delays can lead to the loss of critical information.

\subsection{Fresh2comm Performance Evaluation}
After validating the optimization methods and assessing the impact of delays, we proceed to model real-world delay scenarios. In this section, we compare the effect of our GreedyPA on perception results under various conditions.

It is important to note that we set a perception cycle, \textit{looptime}, and the timestamp of the ground truth is set to the current time plus the \textit{looptime}. Our independent variables include full-scenario communication, as well as single-channel communication with three, four, and five vehicles. 

For comparison, we use the experimental results that do not consider delay as the ideal optimal solution, and the DefaultPA serves as the control group for our method.

\begin{figure}[htbp]
    \centering
    \includegraphics[width=0.9\linewidth]{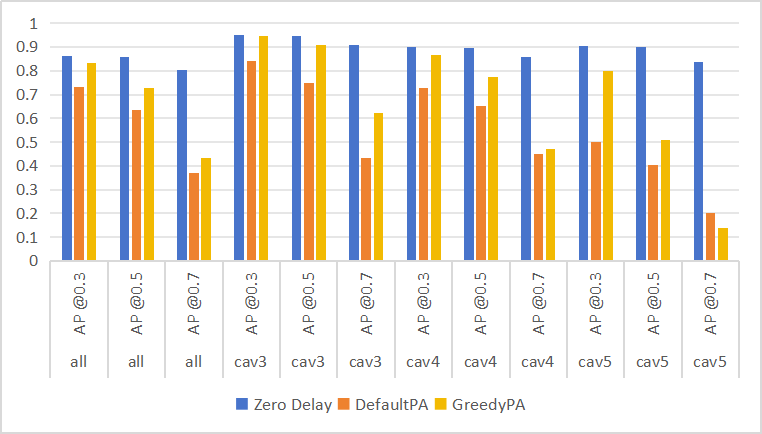}
    \caption{Comparison of Perception Results}
    \label{fig:4.3.1}
\end{figure}

The results are summarized in Table~\ref{tab:performance_results1}. Figure~\ref{fig:4.3.1} illustrates the visualization of the results.

\begin{table}[htbp]
    \centering
    \resizebox{8.5cm}{!}{
    \begin{tabular}{cccccc}
        \toprule
        \textbf{num\_cavs} & \textbf{mode} & \textbf{AP @0.3$\uparrow$} & \textbf{AP @0.5$\uparrow$} & \textbf{AP @0.7$\uparrow$} \\
        \midrule
        all  & Zero Delay   & $0.864$ & $0.860$ & $0.805$ \\
        all  & DefaultPA    & $0.730$ & $0.637$ & $0.370$ \\
        all  & GreedyPA     & $0.834$ & $0.727$ & $0.435$ \\
        \midrule
        cav3 & Zero Delay   & $0.949$ & $0.947$ & $0.910$ \\
        cav3 & DefaultPA    & $0.843$ & $0.748$ & $0.434$ \\
        cav3 & GreedyPA     & $0.947$ & $0.908$ & $0.623$ \\
        \midrule
        cav4 & Zero Delay   & $0.898$ & $0.895$ & $0.858$ \\
        cav4 & DefaultPA    & $0.729$ & $0.654$ & $0.449$ \\
        cav4 & GreedyPA     & $0.865$ & $0.774$ & $0.469$ \\
        \midrule
        cav5 & Zero Delay   & $0.906$ & $0.900$ & $0.835$ \\
        cav5 & DefaultPA    & $0.501$ & $0.404$ & $0.201$ \\
        cav5 & GreedyPA     & $0.799$ & $0.508$ & $0.138$ \\
        \bottomrule
    \end{tabular}
    }
    \caption{Comparison of Perception Results}
    \label{tab:performance_results1}
\end{table}

Since the LiDAR data sampling period is 0.1s, the minimum perception cycle is also 0.1s. Therefore, the data in Table~\ref{tab:performance_results1} serves as the most valuable reference for our baseline. 

By analyzing the data in Table~\ref{tab:performance_results1}, we can first observe that our optimization method achieves significant improvements across the entire dataset compared to the default setting. Although the value of $AP@0.7$ has decreased considerably, the value of $AP@0.3$ remains at a high level. This indicates that, under our optimization method, key information is preserved, and the decrease in accuracy is minimized.

When we compare the results across scenarios involving 3, 4, and 5 vehicles, we find that as the number of vehicles increases, the average delay also increases significantly, leading to a decrease in perception performance for both DefaultPA and GreedyPA. In scenarios with 3 and 4 vehicles, our GreedyPA clearly outperforms DefaultPA across all evaluation metrics. However, in the 5-vehicle scenario, although we achieve a significant improvement in $AP@0.3$, our $AP@0.7$ score falls short of DefaultPA. This is because, while we make efforts to ensure that no information is lost, we sacrifice more in terms of perception accuracy compared to DefaultPA. 

This finding aligns with previous discussions: high variability in AoIs can cause the loss of more information, but the perception accuracy may still be better in scenarios with nonlinear accuracy degradation. The main goal of collaborative perception is to avoid information loss due to field-of-view gaps. Therefore, we believe that in extreme cases, the drop in perception accuracy caused by our optimization method is acceptable.

\begin{table}[htbp]
    \centering
    \resizebox{8.5cm}{!}{
    \begin{tabular}{cccccc}
        \toprule
        \textbf{num\_cavs} & \textbf{mode} & \textbf{AP @0.3$\uparrow$} & \textbf{AP @0.5$\uparrow$} & \textbf{AP @0.7$\uparrow$} \\
        \midrule
        all  & Zero Delay   & $0.864$ & $0.860$ & $0.805$ \\
        all  & DefaultPA    & $0.688$ & $0.537$ & $0.280$ \\
        all  & GreedyPA     & $0.822$ & $0.739$ & $0.395$ \\
        \midrule
        cav3 & Zero Delay   & $0.949$ & $0.947$ & $0.910$ \\
        cav3 & DefaultPA    & $0.842$ & $0.700$ & $0.363$ \\
        cav3 & GreedyPA     & $0.945$ & $0.890$ & $0.458$ \\
        \midrule
        cav4 & Zero Delay   & $0.898$ & $0.895$ & $0.858$ \\
        cav4 & DefaultPA    & $0.710$ & $0.611$ & $0.391$ \\
        cav4 & GreedyPA     & $0.874$ & $0.842$ & $0.658$ \\
        \midrule
        cav5 & Zero Delay   & $0.906$ & $0.900$ & $0.835$ \\
        cav5 & DefaultPA    & $0.418$ & $0.335$ & $0.135$ \\
        cav5 & GreedyPA     & $0.874$ & $0.823$ & $0.404$ \\
        \bottomrule
    \end{tabular}
    }
    \caption{Comparison of Perception Results (looptime=0.2)}
    \label{tab:updated_performance_results_1}
\end{table}

\begin{figure}[htbp]
    \centering
    \includegraphics[width=0.9\linewidth]{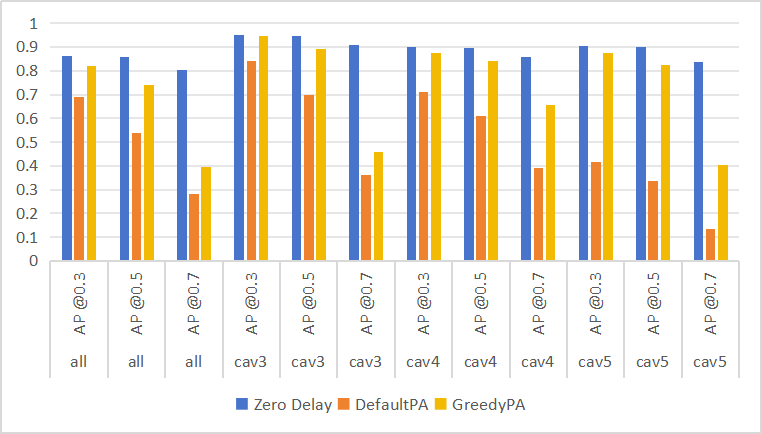}
    \caption{Comparison of Perception Results (looptime=0.2)}
    \label{fig:enter-label2}
\end{figure}

In addition, we conducted two sets of control experiments, as shown in Table~\ref{tab:updated_performance_results_1} and Table~\ref{tab:updated_performance_results_2}. In the experiment result presented in Table~\ref{tab:updated_performance_results_1} and Figure~\ref{fig:enter-label2}, the perception cycle was increased to 0.2s. From the results, we can see that our Greedy Optimization method significantly improves perception performance in scenarios with 4 and 5 vehicles. However, in the 3-vehicle scenario, there is a slight decrease in performance. This indicates that our perception results are sensitive to the choice of perception cycle.

On the other hand, the perception results of DefaultPA degrade uniformly across all scenarios. This suggests that the decline in perception performance due to high variability in AoIs is highly random and difficult to optimize further.

In summary, our optimization method demonstrates considerable potential for improvement across multiple dimensions. For example, the \textit{How2Comm} framework enhances the robustness of perception results by utilizing historical data. This approach would further improve our GreedyPA, but it would not provide additional benefits to DefaultPA.

The third experimental setup is constructed based on the \textit{Where2comm} framework. We calculated the average effective communication rate to be approximately 0.2154, meaning that in an ideal scenario, the transmitted data size could reach 0.2154 times its original size. Building on this, we conducted an additional set of experiments.
\begin{table}[htbp]
    \centering
    \resizebox{8.5cm}{!}{
    \begin{tabular}{cccccc}
        \toprule
        \textbf{num\_cavs} & \textbf{mode} & \textbf{AP @0.3$\uparrow$} & \textbf{AP @0.5$\uparrow$} & \textbf{AP @0.7$\uparrow$} \\
        \midrule
        all  & Zero Delay   & $0.864$ & $0.860$ & $0.805$ \\
        all  & DefaultPA        & $0.766$ & $0.682$ & $0.400$ \\
        all  & GreedyPA     & $0.859$ & $0.802$ & $0.489$ \\
        \midrule
        cav3 & Zero Delay   & $0.949$ & $0.947$ & $0.910$ \\
        cav3 & DefaultPA        & $0.912$ & $0.839$ & $0.507$ \\
        cav3 & GreedyPA     & $0.946$ & $0.897$ & $0.517$ \\
        \midrule
        cav4 & Zero Delay   & $0.898$ & $0.895$ & $0.858$ \\
        cav4 & DefaultPA        & $0.785$ & $0.721$ & $0.510$ \\
        cav4 & GreedyPA     & $0.897$ & $0.866$ & $0.633$ \\
        \midrule
        cav5 & Zero Delay   & $0.906$ & $0.900$ & $0.835$ \\
        cav5 & DefaultPA        & $0.529$ & $0.441$ & $0.218$ \\
        cav5 & GreedyPA     & $0.905$ & $0.875$ & $0.608$ \\
        \bottomrule
    \end{tabular}
    }
    \caption{Comparison of perception Results with Reduced Datasize}
    \label{tab:updated_performance_results_2}
\end{table}

\begin{figure}[htbp]
    \centering
    \includegraphics[width=0.9\linewidth]{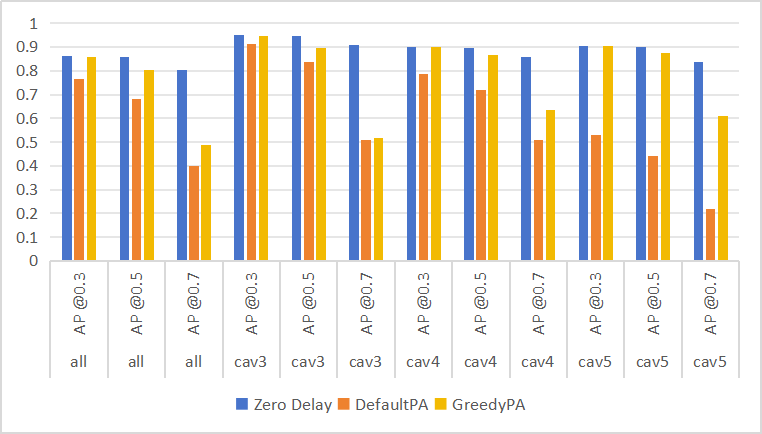}
    \caption{Comparison of perception Results with Reduced Datasize}
    \label{fig:enter-label3}
\end{figure}

The results shown in Table~\ref{tab:updated_performance_results_2} and Figure~\ref{fig:enter-label3} were remarkable. For instance, the $AP@0.7$ score of GreedyPA increased from 0.138 to 0.608, marking a substantial improvement. In comparison, the perception results of DefaultPA also showed some improvement, but the increase was much smaller than that achieved with GreedyPA. This set of data once again demonstrates that our optimization method is highly adaptable for improving other dimensions.

In summary, our Greedy Optimization method significantly reduces communication delays and effectively addresses the issue of high variability in AoIs. This improvement is reflected in the perception results, where our method ensures that information loss is minimized, though some accuracy is inevitably sacrificed.

On the other hand, our method demonstrates strong adaptability in multi-dimensional optimization, making it an ideal candidate for integration into optimization frameworks.

\section{Conclusion}
In conclusion, we construct a communication model based on SOTA perception model that reflecting real-world communication scenarios. Furthermore, the impact of different delay types on perception performance, including computation and communication delays, is critically examined. The findings suggest that minimizing delays and avoiding high variability in AoIs are key factors for optimizing collaborative perception. 

Base on the findings, \textit{Fresh2comm}, a novel AoI-based optimization framework designed to optimize AoI and improve collaborative perception in ICV systems is introduced. The framework is capable of dynamically balancing the preservation of key information and the perception accuracy through advanced optimization method GreedyPA, which provides significant improvements in communication delay reduction and delay variance while maintaining low computational cost. The experimental results validate the effectiveness of GreedyPA in perception accuracy in real-world communication scenarios, while maintaining a high level of information integrity.

\bibliographystyle{named}
\bibliography{main}

\begin{thebibliography}{}

\bibitem[\protect\citeauthoryear{Abdel-Aziz \bgroup \em et al.\egroup }{2020a}]{8918241}
Mohamed~K. Abdel-Aziz, Sumudu Samarakoon, Mehdi Bennis, and Walid Saad.
\newblock Ultra-reliable and low-latency vehicular communication: An active learning approach.
\newblock {\em IEEE Communications Letters}, 24(2):367--370, 2020.

\bibitem[\protect\citeauthoryear{Abdel-Aziz \bgroup \em et al.\egroup }{2020b}]{8937801}
Mohamed~K. Abdel-Aziz, Sumudu Samarakoon, Chen-Feng Liu, Mehdi Bennis, and Walid Saad.
\newblock Optimized age of information tail for ultra-reliable low-latency communications in vehicular networks.
\newblock {\em IEEE Transactions on Communications}, 68(3):1911--1924, 2020.

\bibitem[\protect\citeauthoryear{Alabbasi and Aggarwal}{2022}]{9023385}
Abubakr Alabbasi and Vaneet Aggarwal.
\newblock Joint information freshness and completion time optimization for vehicular networks.
\newblock {\em IEEE Transactions on Services Computing}, 15(2):1118--1129, 2022.

\bibitem[\protect\citeauthoryear{Chen \bgroup \em et al.\egroup }{2020}]{8954939}
Xianfu Chen, Celimuge Wu, Tao Chen, Honggang Zhang, Zhi Liu, Yan Zhang, and Mehdi Bennis.
\newblock Age of information aware radio resource management in vehicular networks: A proactive deep reinforcement learning perspective.
\newblock {\em IEEE Transactions on Wireless Communications}, 19(4):2268--2281, 2020.

\bibitem[\protect\citeauthoryear{Cui \bgroup \em et al.\egroup }{2022}]{DBLP:journals/sensors/CuiZXYF22}
Guangzhen Cui, Weili Zhang, Yanqiu Xiao, Lei Yao, and Zhanpeng Fang.
\newblock Cooperative perception technology of autonomous driving in the internet of vehicles environment: {A} review.
\newblock {\em Sensors}, 22(15):5535, 2022.

\bibitem[\protect\citeauthoryear{Gai \bgroup \em et al.\egroup }{2023}]{10130737}
Hao Gai, Haixia Zhang, Shuaishuai Guo, and Dongfeng Yuan.
\newblock Information freshness-oriented trajectory planning and resource allocation for uav-assisted vehicular networks.
\newblock {\em China Communications}, 20(5):244--262, 2023.

\bibitem[\protect\citeauthoryear{Guo \bgroup \em et al.\egroup }{2022}]{9788527}
Shiqian Guo, Bin-Jie Hu, and Qingji Wen.
\newblock Joint resource allocation and power control for full-duplex v2i communication in high-density vehicular network.
\newblock {\em IEEE Transactions on Wireless Communications}, 21(11):9497--9508, 2022.

\bibitem[\protect\citeauthoryear{Guo \bgroup \em et al.\egroup }{2023}]{9903381}
Chongtao Guo, Xijun Wang, Le~Liang, and Geoffrey~Ye Li.
\newblock Age of information, latency, and reliability in intelligent vehicular networks.
\newblock {\em IEEE Network}, 37(6):109--116, 2023.

\bibitem[\protect\citeauthoryear{Guo \bgroup \em et al.\egroup }{2024}]{10081395}
Chongtao Guo, Songtao Liu, Bin Liao, Zhigang Wang, and Le~Liang.
\newblock Aoi-driven power allocation and batch sampling control for v2v status update communications.
\newblock {\em IEEE Transactions on Industrial Informatics}, 20(1):291--302, 2024.

\bibitem[\protect\citeauthoryear{Hu \bgroup \em et al.\egroup }{2022}]{NEURIPS2022_1f5c5cd0}
Yue Hu, Shaoheng Fang, Zixing Lei, Yiqi Zhong, and Siheng Chen.
\newblock Where2comm: Communication-efficient collaborative perception via spatial confidence maps.
\newblock In S.~Koyejo, S.~Mohamed, A.~Agarwal, D.~Belgrave, K.~Cho, and A.~Oh, editors, {\em Advances in Neural Information Processing Systems}, volume~35, pages 4874--4886. Curran Associates, Inc., 2022.

\bibitem[\protect\citeauthoryear{Hu \bgroup \em et al.\egroup }{2024}]{hu2024collaborativeperceptionconnectedautonomous}
Senkang Hu, Zhengru Fang, Yiqin Deng, Xianhao Chen, and Yuguang Fang.
\newblock Collaborative perception for connected and autonomous driving: Challenges, possible solutions and opportunities, 2024.

\bibitem[\protect\citeauthoryear{Kaul \bgroup \em et al.\egroup }{2012}]{6195689}
Sanjit Kaul, Roy Yates, and Marco Gruteser.
\newblock Real-time status: How often should one update?
\newblock In {\em 2012 Proceedings IEEE INFOCOM}, pages 2731--2735, 2012.

\bibitem[\protect\citeauthoryear{Liu \bgroup \em et al.\egroup }{2020}]{9156848}
Yen-Cheng Liu, Junjiao Tian, Nathaniel Glaser, and Zsolt Kira.
\newblock When2com: Multi-agent perception via communication graph grouping.
\newblock In {\em 2020 IEEE/CVF Conference on Computer Vision and Pattern Recognition (CVPR)}, pages 4105--4114, 2020.

\bibitem[\protect\citeauthoryear{Sorkhoh \bgroup \em et al.\egroup }{2022}]{9599398}
Ibrahim Sorkhoh, Chadi Assi, Dariush Ebrahimi, and Sanaa Sharafeddine.
\newblock Optimizing information freshness for mec-enabled cooperative autonomous driving.
\newblock {\em IEEE Transactions on Intelligent Transportation Systems}, 23(8):13127--13140, 2022.

\bibitem[\protect\citeauthoryear{Xu \bgroup \em et al.\egroup }{2022}]{10.1109/ICRA46639.2022.9812038}
Runsheng Xu, Hao Xiang, Xin Xia, Xu~Han, Jinlong Li, and Jiaqi Ma.
\newblock Opv2v: An open benchmark dataset and fusion pipeline for perception with vehicle-to-vehicle communication.
\newblock In {\em 2022 International Conference on Robotics and Automation (ICRA)}, page 2583–2589. IEEE Press, 2022.

\bibitem[\protect\citeauthoryear{Yang \bgroup \em et al.\egroup }{2023a}]{NEURIPS2023_4f31327e}
Dingkang Yang, Kun Yang, Yuzheng Wang, Jing Liu, Zhi Xu, Rongbin Yin, Peng Zhai, and Lihua Zhang.
\newblock How2comm: Communication-efficient and collaboration-pragmatic multi-agent perception.
\newblock In A.~Oh, T.~Naumann, A.~Globerson, K.~Saenko, M.~Hardt, and S.~Levine, editors, {\em Advances in Neural Information Processing Systems}, volume~36, pages 25151--25164. Curran Associates, Inc., 2023.

\bibitem[\protect\citeauthoryear{Yang \bgroup \em et al.\egroup }{2023b}]{10.1145/3581783.3611699}
Kun Yang, Dingkang Yang, Jingyu Zhang, Hanqi Wang, Peng Sun, and Liang Song.
\newblock What2comm: Towards communication-efficient collaborative perception via feature decoupling.
\newblock In {\em Proceedings of the 31st ACM International Conference on Multimedia}, MM '23, page 7686–7695, New York, NY, USA, 2023. Association for Computing Machinery.

\bibitem[\protect\citeauthoryear{Yu \bgroup \em et al.\egroup }{2021}]{9709015}
Ruozhou Yu, Dejun Yang, and Hao Zhang.
\newblock Edge-assisted collaborative perception in autonomous driving: A reflection on communication design.
\newblock In {\em 2021 IEEE/ACM Symposium on Edge Computing (SEC)}, pages 371--375, 2021.

\bibitem[\protect\citeauthoryear{Yu \bgroup \em et al.\egroup }{2022}]{9879243}
Haibao Yu, Yizhen Luo, Mao Shu, Yiyi Huo, Zebang Yang, Yifeng Shi, Zhenglong Guo, Hanyu Li, Xing Hu, Jirui Yuan, and Zaiqing Nie.
\newblock Dair-v2x: A large-scale dataset for vehicle-infrastructure cooperative 3d object detection.
\newblock In {\em 2022 IEEE/CVF Conference on Computer Vision and Pattern Recognition (CVPR)}, pages 21329--21338, 2022.

\end{thebibliography}

\end{document}